\documentstyle[12pt]{article}
\begin{document}
\date{}
\title{The Finite Temperature Effective Potential for Local Composite Operators}
\author{Anna Okopi\'nska\\ 
Institute of Physics, Warsaw University, Bia\l ystok Branch,\\ 
Lipowa 41, 15-424 Bia\l ystok, Poland\\e-mail: rozynek@fuw.edu.pl}
\maketitle
\thispagestyle{empty}
\begin{abstract}
\noindent 
The method of the effective action for the composite operators
$\Phi^2(x)$ and $\Phi^4(x)$ is applied to the termodynamics of
the scalar quantum field with $\lambda\Phi^4$ interaction. An
expansion of the finite temperature effective potential in
powers of $\hbar$ provides successive approximations to the free
energy with an effective mass and an effective coupling
determined by the gap equations. The numerical results are
studied in the space-time of one dimension, when the theory is
equivalent to the quantum mechanics of an anharmonic oscillator.
The approximations to the free energy show quick convergence to
the exact result.
\end{abstract}
\newpage
\section{Introduction}

Thermal properties of quantum field systems are important for a
study of phase transitions in the big bang cosmology and for an
interpretation of heavy ion collisions. These systems are
characterised by a large coupling constant, thus
non-perturbative methods should be developped to study their
termodynamics. Here we shall discuss an application of the
effective action for local composite operators~\cite{Fukinv} to
the finite temperaturequantum field theory. We consider a scalar field with a
classical action in $n$-dimensional Euclidean space-time given
by
\begin{equation}
S[\Phi]=\int\![\frac{1}{2}\Phi(x)(-\partial^2+m^2)\Phi(x)+
\lambda\Phi^{4}(x)]\,d^{n}x,
\label{Scl}
\end{equation}
but the method can be easily extended to fermionic systems.

Recently, the diagrammatic rules for the effective action for
the operator $\Phi^2(x)$ have been found~\cite{OY} and the
method has been extended to include the operator
$\Phi^4(x)$~\cite{AOinv}. The lowest order approximation
coincides with the Gaussian Effective Action resulting from the
variational principle with Gaussian trial states~\cite{GEP}.
Within the thermal field theory formalism this would correspond
to the finite temperature Hartree approximation for the free
energy~\cite{Rim}. An expansion of the effective action for
local composite operators in powers of $\hbar$ provides thus a
non-perturbative method for a systematic improvement of the
Hartree approximation both at $T=0$ and $T\ne 0$. An advantage
of using local operators lies in the fact that calculation is
simpler than for bilocal composite operators
$\Phi(x)\Phi(y)$~\cite{CJT} and other systematic methods which
contain the Gaussian Effective Action as the lowest
approximation (f.e. the optimized expansion~\cite{AO}).

\section{Finite temperature effective potential for composite operators}
\label{inv}
The vacuum functional for the composite operators $\Phi^2(x)$
and $\Phi^4(x)$ may be represented by a path integral
\begin{equation}
Z[J,K]=e^{\frac{1}{\hbar}W[J,K]}=\int\!
D\Phi\,e^{\frac{1}{\hbar}\left[-S[\Phi] +
\frac{1}{2}\int\!\!J(x)\Phi^{2}(x)\,d^{n}x+
\frac{1}{24}\int\!\!K(x)\Phi^{4}(x)\,d^{n}x\right]}
\label{Z4}
\end{equation}
and the effective action is defined through a Legendre transform
\begin{eqnarray}
\Gamma[\Delta,\Lambda]&=&W[J,K]-\frac{\hbar}{2}\int\! J(x)
\Delta(x)\,d^{n}x -\frac{\hbar^3}{24}\int\! K(x)
\Lambda(x)\,d^{n}x\nonumber\\
&-&\frac{\hbar^2}{8}\int\! K(x)\Delta^2(x)\,d^{n}x,
\label{Gam}
\end{eqnarray}
where the background fields are given by
\begin{eqnarray}
\hbar\Delta(x)&=&2 \frac{\delta W}{\delta J(x)}=<\Phi^2(x)>_{J,K}\nonumber\\
\hbar^3\Lambda(x)&=&24 \frac{\delta W}{\delta K(x)}-3\hbar^2\Delta^2(x)=
<\Phi^4(x)>_{J,K}-3<\Phi^2(x)>_{J,K}^2
\label{dJK}
\end{eqnarray}
and $<...>_{J,K}$ denotes the expectation value in the presence
of external sources $J$ and $K$. The physical quantities have to
be calculated at {\mbox{$J=K=0$}}, or equivalently at the values of the
background fields for which the gap equations
\begin{eqnarray}
\frac{\delta\Gamma}{\delta \Delta(x)}&=&-\frac{\hbar}{2}J(x)-
\frac{\hbar^2}{4}\Delta(x) K(x)=0,\nonumber\\
\frac{\delta\Gamma}{\delta \Lambda(x)}&=&-\frac{\hbar^3}{24}K(x)=0
\label{gap}
\end{eqnarray}
are satisfied.
 
The path-integral quantization in Euclidean space-time enables
us to study the quantum field theory and its equilibrium
thermodynamics by the same "imaginary time" formalism, if the
appropriate boundary conditions are chosen~\cite{ITF,DJ}. The
vacuum functional~(\ref{Z4}) defines the generating functional
for composite Green's functions in quantum field theory, if the
path integral is taken over the functions which approaches some
constant as $x_{0}\rightarrow\pm\infty$. If the set of functions
is periodic with a period $\beta=\frac{1}{T}$ in the "imaginary
time coordinate" $x_{0}$, the integral~(\ref{Z4}) has a meaning
of the partition function at the temperature $T$, generalised to
include interactions with the external currents $J$ and $K$. The
finite temperature effective potential for the composite
operators $\Phi^2(x)$ and $\Phi^4(x)$ is defined as
\begin{equation}
V_{\beta}(\Delta,\Lambda)=-\left.\frac{\Gamma_{\beta}[\Delta,\Lambda]}
{\beta\int\!d^{n-1}x}\right|_{\Delta=const,\Lambda=const}.
\label{FTEP}
\end{equation} 
The free energy density can be obtained as
$F_{\beta}=V_{\beta}(\Delta_{\beta},\Lambda_{\beta})$, where the
temperature dependent expectation values of the composite
fields, $\Delta_{\beta}$ and $\Lambda_{\beta}$, which correspond
to $J=K=0$, are determined by a stationary point of
$V_{\beta}(\Delta,\Lambda)$, according to~(\ref{gap}).

Representing the connected generating functional by the series
\begin{equation} 
W[J,K]=\sum_{k=0}^{\infty}\hbar^k W_{(k)}[J,K],
\label{WW}
\end{equation}
the effective action for composite operators~(\ref{Gam}) may be
obtained by eliminating the sources
\begin{equation} 
J=\sum_{k=0}^{\infty}\hbar^k J_{(k)}[\Delta,\Lambda],
\mbox{~~~~~and~~~~~}
K=\sum_{k=0}^{\infty}\hbar^k K_{(k)}[\Delta,\Lambda]
\label{dlap}
\end{equation} 
in favour of the expectation values of the composite fields
\begin{equation} 
\Delta=\sum_{k=0}^{\infty}\hbar^k \Delta_{(k)}[J,K],
\mbox{~~~~~and~~~~~}
\Lambda=\sum_{k=0}^{\infty}\hbar^k \Lambda_{(k)}[J,K],
\label{dKp}
\end{equation}
order by order in $\hbar$. The coefficients $J_{0}$ and $K_{0}$
are determined implicitly by the lowest-order relations
\begin{equation}
\unitlength=1.00mm
\linethickness{0.4pt}
\begin{picture}(141.00,3.00)
\put(72.00,-3.00){\circle{12.}}
\put(66.00,-3.00){\circle{2.00}}
\put(7.00,-3.00){\makebox(0,0)[lc]{$\Delta(x)=\Delta_{0}(x)=G(x,x)=~(x)$}}
\end{picture}
\label{d0}
\end{equation}
\\
\noindent and
\begin{equation}
\unitlength=1.00mm
\linethickness{0.4pt}
\begin{picture}(141.00,3.00)
\put(7.00,-4.00){\makebox(0,0)[lc]{
$\Lambda(x)=\Lambda_{0}(x)=\int G^{4}(x,y)(K_{0}(y)-24\lambda)d^{n}y,$}}
\end{picture}
\label{l0}
\end{equation}
\\
\noindent where the inverse propagator is given by
\begin{equation}
G^{-1}(x,y)=(-\partial ^2+\Omega^2(x))\delta(x-y),
\label{prop}
\end{equation}
\noindent with an effective mass $\Omega(x)$ defined by
\begin{equation}
\Omega^2[\Delta]=m^2-J_{0}[\Delta].
\label{om}
\end{equation}
Higher coefficients, $J_{(k)}$ and $K_{(k)}$ may be easily
expressed as functionals of $J_{0}[\Delta]$ and
$K_{0}[\Delta,\Lambda]$ and the Legendre transform~(\ref{Gam})
can be performed. To the order $\hbar^4$ the result is~\cite{AOinv}
\begin{equation}
\unitlength=1.00mm
\linethickness{0.4pt}
\begin{picture}(120.00,29.00)
\put(53.00,1.00){\circle{12.00}}
\put(78.00,1.00){\circle{12.00}}
\put(62.00,1.00){\makebox(0,0)[lc]{$-\frac{\hbar^3}{48}$}}
\put(86.00,1.00){\makebox(0,0)[lc]{$+\frac{\hbar^4}{48}$}}
\put(102.00,1.00){\circle{12.00}}
\put(78.00,1.00){\oval(12.00,6.00)[]}
\put(97.00,-3.00){\line(1,2){5.00}}
\put(107.00,-3.00){\line(-1,2){5.00}}
\put(97.00,-3.00){\line(1,0){10.00}}
\put(7.00,20.00){\makebox(0,0)[lc]{$\Gamma[\Delta,\Lambda]=\int[\frac{\hbar}{2}(\Omega^2(x)-m^2)\Delta(x)-3\lambda\hbar^2\Delta^2(x)-\hbar^3\lambda\Lambda(x)]dx$}}
\put(37.00,1.00){\makebox(0,0)[lc]{$-\frac{\hbar}{2}$}}
\put(72.00,1.00){\circle*{2.00}}
\put(84.00,1.00){\circle*{2.00}}
\put(97.00,-3.00){\circle*{2.00}}
\put(107.00,-3.00){\circle*{2.00}}
\put(102.00,7.00){\circle*{2.00}}
\put(113.00,1.00){\makebox(0,0)[lc]{$+~...$}}
\end{picture}
\label{Gam4}
\end{equation}
\\
with $\Omega$ related to $\Delta$ by~(\ref{d0}) and (\ref{prop}), 
and the effective coupling 
\begin{equation}
\unitlength=1.00mm
\linethickness{0.4pt}
\begin{picture}(141.00,6.00)
\put(10.00,-2.00){\circle*{2.00}}
\put(8,-4){\line(1,1){4}}
\put(8,0){\line(1,-1){4}}
\put(20.00,-2.00){\makebox(0,0)[lc]{$=$}}
\put(34.00,-2.){\oval(12.00,6.00)[]}
\put(34.00,-2.){\circle{12.00}}
\put(28.00,-2.00){\circle{2.00}}
\put(28,-8){\line(1,1){12}}
\put(42.00,-2.00){\makebox(0,0)[lc]{$\Lambda,$}}
\end{picture}
\label{l00}
\end{equation}
\\
\noindent where a slash denotes an inversion of the operator.

The effective action which includes the two-particle operator
$\Phi^2(x)$ only, $\Gamma[\Delta]$, can be obtained from the
above expression for $\Gamma[\Delta,\Lambda]$ by setting
$K_{0}=0$ which results in 
\begin{equation}
\unitlength=1.00mm
\linethickness{0.4pt}
\begin{picture}(120.00,25.00)
\put(53.00,1.00){\circle{12.00}}
\put(78.00,1.00){\circle{12.00}}
\put(62.00,1.00){\makebox(0,0)[lc]{$+\frac{\hbar^3}{48}$}}
\put(86.00,1.00){\makebox(0,0)[lc]{$+\frac{\hbar^4}{48}$}}
\put(102.00,1.00){\circle{12.00}}
\put(78.00,1.00){\oval(12.00,6.00)[]}
\put(97.00,-3.00){\line(1,2){5.00}}
\put(107.00,-3.00){\line(-1,2){5.00}}
\put(97.00,-3.00){\line(1,0){10.00}}
\put(7.00,20.00){\makebox(0,0)[lc]{$\Gamma[\Delta]=\int[\frac{\hbar}{2}(\Omega^2(x)-m^2)\Delta(x)-3\lambda\hbar^2\Delta^2(x)]dx$}}
\put(37.00,1.00){\makebox(0,0)[lc]{$-\frac{\hbar}{2}$}}
\put(113.00,1.00){\makebox(0,0)[lc]{$+~...$}}
\end{picture}
\label{Gam2}
\end{equation}
\\
in agreement with diagrammatic rules proven in Ref.~\cite{OY}.

The expansions of $\Gamma[\Delta]$ and $\Gamma[\Delta,\Lambda]$
in powers of $\hbar$ provide two different approximation schemes
for physical quantities, we shall denote these as I2 and I4,
respectively. The non-perturbative character of the
approximations is due to the fact that the expectation values of
the composite fields are determined from the gap equations,
obtained by requiring the given order expression for the
effective action to be stationary. At order $\hbar^2$ both
approaches results in the Gaussian effective action~\cite{GEP},
which contains an infinite sum of bubble diagrams of
perturbation theory. As it happens, the approximations to
physical quantities obtained from $\Gamma[\Delta]$ and
$\Gamma[\Delta,\Lambda]$ coincide also at order $\hbar^3$, after
solving the gap equations, and differences first appear at order
$\hbar^4$. One can note that by introducing higher composite
operators new approximation schemes can be formulated, but their
influence will appear in much higher orders of $\hbar$.

The finite temperature effective potential for composite
operators can be obtained from the effective action by replacing
the Feynman rules by those at finite temperature. In the given
order approximation to the effective action the infinite
interval in all integrals over $t$ is replaced by $[0,\beta]$.
$\Delta$ and $\Lambda$ are taken to be space-time independent,
in this case $\Omega$ is also constant and the Fourier transform
of the propagator~(\ref{prop}) can be performed. Using a
momentum representation in space, but a coordinate
representation in time we have
\begin{equation}
G_{\beta}(\tau,{\bf p})=\frac{1}{\beta}\sum_{m=-\infty}^{\infty}
\frac{\exp[-i\omega_{m}(\tau)]}{{\bf p}^2+\omega_{m}^{2}+\Omega^2}
\label{Gbeta}
\end{equation}
where $\omega_{m}=\frac{2\pi m}{\beta}$ are the Matsubara
frequencies. The free energy calculated at the order $\hbar^2$,
appears identical with that in the Hartree approximation,
provided by the finite temperature Gaussian Effective
Potential~\cite{Rim}. The post-Gaussian corrections
can be determined by taking into account higher orders of
the finite temperature effective potential for composite
operators.

\section{Free energy of the quantum-mechanical anharmonic
\mbox{oscillator}}
\label{QM}
In the space-time of one dimension (time) the $\lambda\Phi^4$
theory is equivalent to a quantum mechanical anharmonic
oscillator with a Hamiltonian 
\begin{equation}
H=\frac{1}{2} p^2+\frac{1}{2} m^2 x^2+\lambda x^4.
\label{AO}
\end{equation}
After rescaling all quantities in terms of $\lambda$, only one
dimensionless parameter $z=\frac{m^2}{2\lambda^{2/3}}$ remains;
therefore, when discussing numerical results, we put $\lambda=1$
without a loss of generality. The spectrum of the anharmonic
oscillator can be calculated numerically and provides the
simplest test for approximation methods in quantum field theory.
The free energy, which contains information on the whole
spectrum, can be used to study the reliability of
field-theoretical methods at finite temperature. The
conventional loop expansion for the free energy coincides with
termodynamic perturbation theory~\cite{Sch}, the
non-perturbative methods provide better approximations.

Approximations to the ground state energy and to the second and
fourth excitation of the anharmonic oscillator, calculated with
the use of operators $\Phi^2(x)$ and $\Phi^4(x)$ have been
shown to converge quickly to the exact results~\cite{AOinv}; we
shall discuss therefore the approximations to the free energy
calculated from the termal effective potential for these
operators. The temperature dependent propagator~(\ref{Gbeta}) in
one dimensional space-time becomes equal to
\begin{equation}
G_{\beta}(\tau)=\frac{\cosh[\frac{\Omega}{2}(|\tau|-\beta)]}
{\Omega \sinh[\frac{\beta\Omega}{2}]}.
\label{Gbeta1}
\end{equation}
and the integrals over $\beta\ge t\ge 0$ in Feynman
diagrams can be easily performed. We shall compare two
approximation methods:\\
(I2) using the operator $\Phi^2(x)$:
\begin{quote}
$F_{\beta}=V_{\beta}(\Delta_{\beta})$, where $V_{\beta}(\Delta)$
is calculated from Eq.~(\ref{Gam2}) and $\Delta_{\beta}$ is a
solution of the gap equation
\begin{eqnarray}
\frac{\delta V_{\beta}}{\delta \Delta(x)}=0
\label{gap2}
\end{eqnarray}
\end{quote}
(I4) using the operators $\Phi^2(x)$ and $\Phi^4(x)$:\\
\begin{quote}
$F_{\beta}=V_{\beta}(\Delta_{\beta},\Lambda_{\beta})$, where
$V_{\beta}(\Delta,\Lambda)$ is calculated from~\ref{Gam4}
and $\Delta_{\beta}$ and $\Lambda_{\beta}$ are obtained as a
solution of the gap equations
\begin{eqnarray}
\frac{\delta V_{\beta}}{\delta \Delta(x)}=0 \mbox{~~~~and~~~~}
\frac{\delta V_{\beta}}{\delta \Lambda(x)}=0.
\label{gap4}
\end{eqnarray}
\end{quote}
The non-perturbative character of the
approximations to the free energy is kept by solving the gap
equations non-perturbatively (if one expanded the solutions of
the gap equations to the given order in $\hbar$, the results
would coincide to that order in $\lambda$ with that obtained in
thermodynamic perturbation theory). In both cases we choose the
solution of the gap equations with the largest positive value
for $\Omega$. The free energy for three values of the parameter
$z=\frac{m^2}{2\lambda^{2/3}}$, $z=5,0,-1$, is shown as a
function of the inverse temperature $\beta=\frac{1}{T}$ in
Figs.~(1)-(3), respectively. The results obtained from
$V_{\beta}(\Delta)$ in tree lowest orders of $\hbar$ (I21, I22
and I23) are compared with that obtained from
$V_{\beta}(\Delta,\Lambda)$ at order $\hbar^3$ (I43). The exact
free energy was obtained as $F_{\beta}=-\frac{1}{\beta}\ln
\sum_{n}e^{-\beta E_{n}}$ where the energy levels, $E_{n}$, have
been calculated by a linear variational method, using harmonic
oscillator wave functions with an appropriately chosen
frequency~\cite{AOAO}.

The comparison shows that an expansion of the termal effective
potential for composite operators in powers of $\hbar$ provides
a method to improve the finite temperature Hartree approximation
(I21) in a systematic way. Both the methods, I2 and I4, work
very well for the single well anharmonic oscillator, the last
being better in the whole range of temperatures. In both cases a
quality of approximations becomes worse for decreasing values of
$z$ and in the double well case ($z~<~0$) the methods are
applicable only if the wells are not too deep ($z~>~-1$). For
$z=-1$ (Figure 3) large discrepancies between the results of
different orders of the method I2 and the exact value appear.
The results obtained at order $\hbar^3$ with the use of the
operators $\Phi^2(x)$ and $\Phi^4(x)$ (I43) are definitely
better than that obtained with the operator $\Phi^2(x)$ only
(I23). At low temperature discrepancies do not exceed 2\%,
but they grow with increasing temperature, where a contibution
of higher excitations becomes more important. This can be
understood from the analysis of the method I4 for the spectrum
of the anharmonic oscillator~\cite{AOinv}, where it has been
demonstated that the quality of the approximations becomes worse
and worse for increasing excitation level. There is a hope that
including higher composite operators will broaden the range of a
good convergence further, making the method applicable even in
the case of the double well potential with deeper wells.
It will be also interesting to discuss an application of the
local composite operators method to study thermodynamics of the scalar
quantum field theory in the case of untrivial space dimension,
we reserve this matter for a future publication.\\

\noindent {\Large {\bf Acknowledgement}}\\

\noindent This work has been supported in part by Grant
2-P03B-048-12 of the Committee for Scientific Research.

\newpage
\noindent {\Large {\bf Figure captions}}\\

\noindent Figure 1. The free energy of the anharmonic oscillator
at $z=\frac{m^2}{2 \lambda^{\frac {2}{3}}}=5$, plotted
{\it vs} $\beta=\frac{1}{T}$. The successive orders
approximations obtained from $V_{\beta}[\Delta]$ ({\it dashed
line}, I21, I22, I23) and from $V_{\beta}[\Delta,\Lambda]$ ({\it
dotted line}, I43) compared with the exact value ({\it solid line}).\\

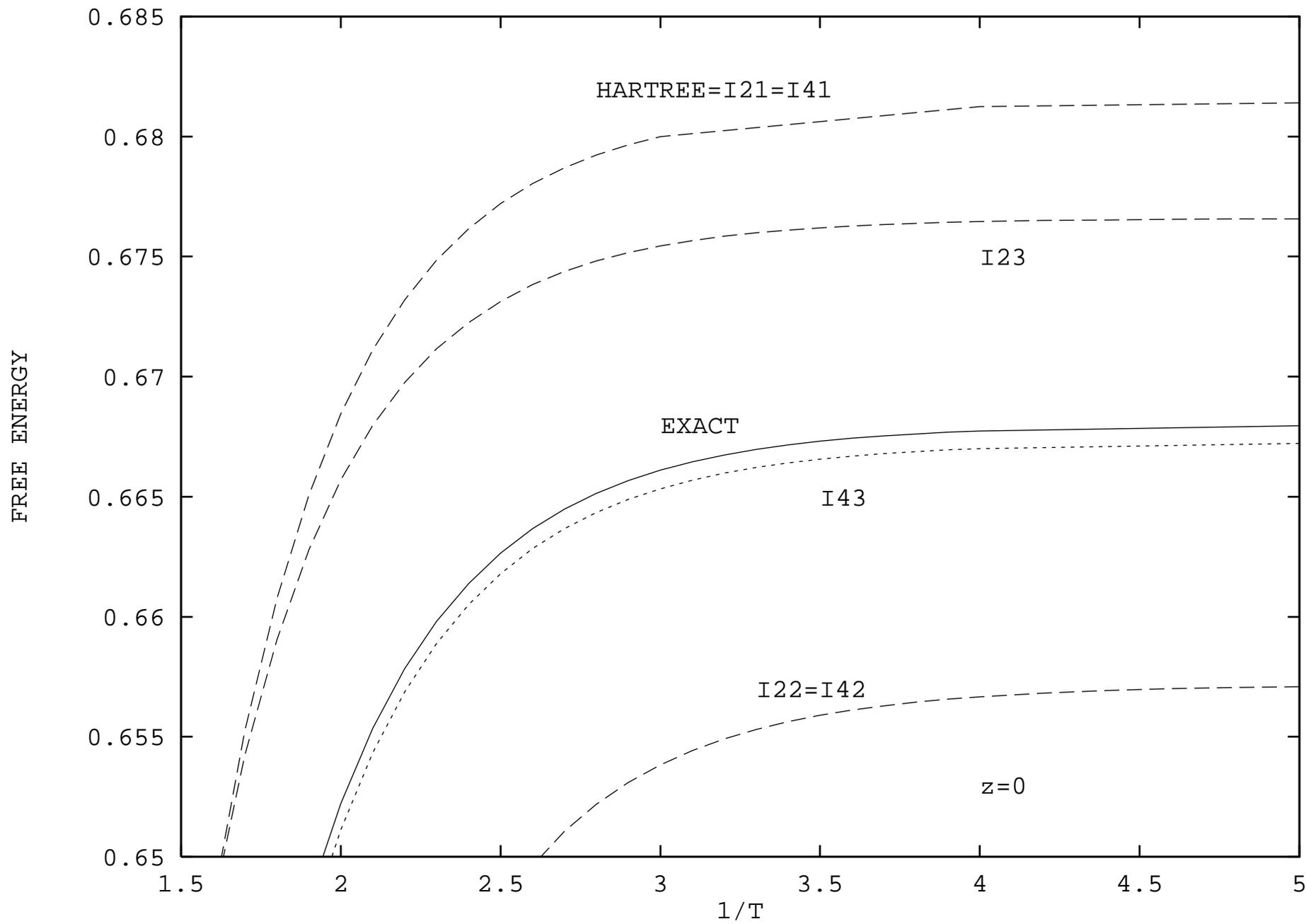
\noindent Figure 2. As in Fig.1, but at $z=0$.\\

\noindent Figure 3. As in Fig.1, but at $z=-1$.\\
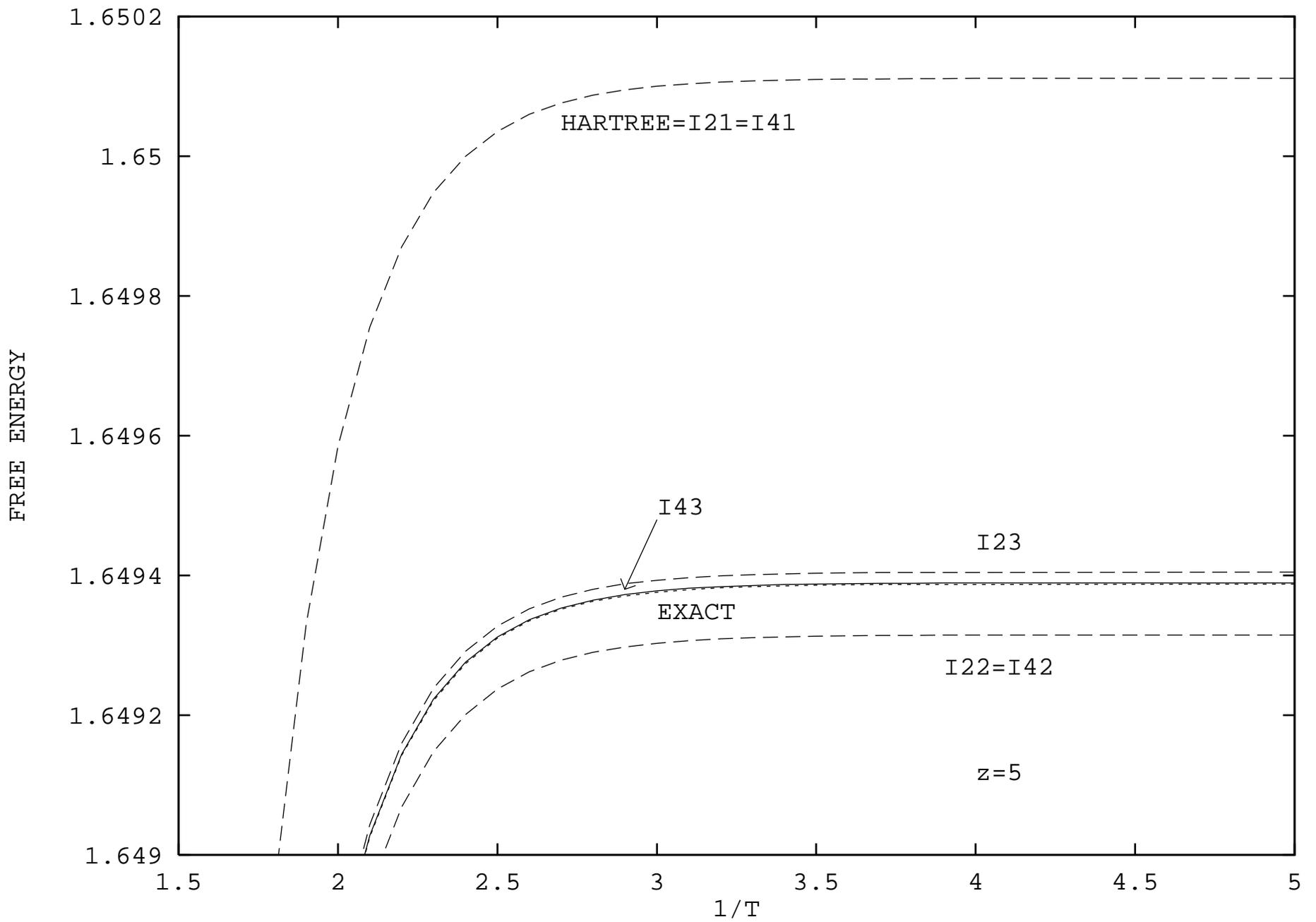
\end{document}